\RequirePackage{tikz}
\documentclass[sn-mathphys]{sn-jnl}

\jyear{2022}%

\usepackage{amssymb,amsfonts}
\usetikzlibrary{arrows}
\usetikzlibrary{positioning}
\usetikzlibrary{fit}
\usepackage{amsthm,amsmath}

\usepackage{lineno}

\usepackage{multirow}

\renewcommand{\epsilon}{\varepsilon}

\begin{document}


\title[A stochastic vector-borne household model]{A stochastic household model for vector-borne diseases}

\author[1]{\fnm{Andrew} \sur{Black}}
\author[2]{\fnm{Andrew} \sur{Smith}}
\author[3]{\fnm{Alun} \sur{Lloyd}}
\author*[1]{\fnm{Joshua} \sur{Ross}}\email{joshua.ross@adelaide.edu.au}

\affil*[1]{\orgdiv{School of Mathematical Sciences}, \orgname{The University of Adelaide}, \orgaddress{\city{Adelaide}, \state{SA}, \postcode{5005},  \country{Australia}}}

\affil[2]{\orgname{Australian Bureau of Statistics}, \orgaddress{\city{Brisbane}, \state{QLD}, \postcode{4000},  \country{Australia}}}

\affil[3]{\orgdiv{Department of Mathematics}, \orgname{North Carolina State University}, \orgaddress{\city{Raleigh}, \state{NC}, \postcode{27695-8213}, \country{United States of America}}}

\abstract{
We introduce a stochastic household model for vector-borne diseases, in particular as relevant to prominent vectors belonging to the {\em Aedes} genus and hence the Zika, chikungunya, and dengue viruses. In this model, vectors remain local to each household, while hosts mix for a proportion of their time in their household and the remaining proportion in the population at random. This is approximated with a two-type branching process, allowing us to efficiently calculate a number of useful epidemiological characteristics, such as reproductive numbers, early growth rates and household-type proportions, offspring distributions, probabilities of a major outbreak, and within-household final size distributions. We compare control interventions of spraying -- reducing the number of vectors in each household -- and social-distancing -- having individuals spend more time at home -- in terms of these characteristics. }


\maketitle



\section{Introduction}
\label{sec:introduction}

Vector-borne diseases, including chikungunya, dengue and Zika, produce significant global burden. Dengue alone is estimated to afflict 104-million people resulting in 40,000 deaths each year~\cite{Zeng2021}. As such there is much interest in understanding the dynamics of these diseases, the potential for their control, and quantifying the impacts of potential interventions.

It has been observed that human dengue incidence and seroprevalence display short-term spatial clustering~\cite{Getis2003, Jeffery2009, Anders2015}. A major contributor to this has been identified, as the main vector {\em Aedes aegypti} are highly anthropophilic, clustering around households, and display restricted dispersal and frequent blood feeding~\cite{Harrington2001, Getis2003, Stoddard2013, Anders2015}. A consequence is that human movement must play a key role in the spread of diseases for which {\em Ae.~aegypti} is the dominant vector.

A number of mathematical models have been developed and analysed in the literature that have captured host-vector disease dynamics and human movement~\cite{Adams2009, Stevens2013, Perkins2013, Reiner2014, Cosner2015, Schaber2021, Tocto-Erazo2021}. The majority of these have focused on deterministic models consisting of a few locations or in continuous space. Only a couple of papers have incorporated stochasticity and household structure explicitly; these models are detailed, simulation-based models~\cite{Reiner2014, Schaber2021}. These models have highlighted the importance of household structure and host movement in regard to understanding vector-borne disease dynamics. Here we extend stochastic household models (i.e., two-level mixing models)~\cite{Ball:1997}, which have found much use in modelling directly-transmitted diseases due to their analytical tractability~\cite{RHK10,BHKR12,Ross2015}, to the situation of vector-borne diseases.

The model introduced herein incorporates two levels of mixing -- within-households and between-households -- and vector-host dynamics. Pertinently, the modelling framework is highly appropriate as the numbers of both humans and mosquitoes at the household-level are small, with estimates of mosquito abundance ranging from three to 40~\cite{Getis2003,Jeffery2009}. Other important features are that the model explicitly accounts for local depletion of susceptible humans within the household, and the natural turnover of vectors due to births and deaths.

We can exploit and extend an extensive body of methodology that has been developed for such household models in the recent literature, allowing us to efficiently calculate: (i) next generation matrices of reproduction numbers, and consequently the household basic reproduction number, $R_*$, and the expected proportion of household types during the exponentially-growing phase of an outbreak; (ii) the early growth rate, $r$, of an outbreak; (iii) the probability mass functions of new households infected over the lifetime of virus presence within a household (i.e., the so-called offspring distribution); (iv) the probabilities of a major outbreak, starting with virus presence within a household; and, (v) the probability mass functions of the number of hosts and vectors within a household infected starting with virus presence within the household (i.e., the so-called final size distributions). 

As effectively a branching process formulation of the problem, our model is most appropriate for the early stages of an outbreak, or for consideration of the potential for an outbreak, and subsequent early dynamics, following introduction of a virus to a household-structured community. We hence analyse two control options: (i) the first reducing the number of vectors in each household, for example through spraying; and, (ii) enacting social-isolation measures of various degrees (from advice to minimise movements for a period to enforced stay-at-home-orders) which increases the proportion of time spent in ones own household, on the probabilities of a major outbreak, and the household basic reproduction number~\cite{Ball:1997,Diekmann2000,RHK10,BHKR12,Ross2015}.

\section{Model}
\label{sec:model}

We adopt a stochastic household modelling framework with two levels of mixing \cite{Ball:1997,RHK10,BHKR12}.  
We first introduce the within-household model of transmission, before detailing how we model the spread between households and how epidemiological quantities of interest can be calculated. 
We assume that each household has a fixed population of hosts of number $N$, and vectors of number $M$. Hosts mix within the overall population, but vectors remain within their household. Hosts are modelled as being either susceptible, infected or recovered; the mosquitoes additionally have an exposed period and birth/ death dynamics linked at the individual level (i.e., SEIS dynamics)~\cite{Black2009}, in which it is assumed that when a mosquito dies it is instantaneously replaced by a susceptible mosquito (so the number of vectors remains constant). 
 
The within-household process is modelled as a continuous-time Markov chain (CTMC), where we let $X$, $Y$ and $Z$ denote the number of susceptible, infected and recovered hosts, respectively, within a house, and $S$, $E$, and $I$ the number of susceptible, exposed and infected vectors, respectively. As $N$ and $M$ are fixed, the state of the process can be specified by
\begin{equation}
\Psi(t) = \left( X(t), Y(t), S(t), E(t)\right),
\end{equation}
where $Z= N-X-Y$, and $I=M-S-E$. The state space for within-household dynamics is then
\begin{equation}
\mathbb{S} = \{ (X, Y, S, E) \, 0 \le X,Y,S,E,\, \, X+Y \le N,\, S+E \le M\}.
\end{equation}

We make the common assumption for host-vector models, that transmission is proportional to the numbers of mosquitoes but the (household) density of the human population \cite{Ross1910,Macdonald1957}. This follows as the number of bites by mosquitoes (so potential transmission events) is proportional to the number of mosquitoes, but independent of the number of humans. 
We decompose the transmission rate into the product of three factors: (i) a common rate of biting per vector within the household, $\beta$; (ii) the probability of transmission upon biting, which is potentially asymmetric capturing that the probability of transmission from vector to host, $p$, or host to vector, $q$, may be different; and, (iii) the proportion of time spent within the primary household, $\alpha$ (and complementary proportion spent visiting other households, $1-\alpha$), where we assume $0.5 > \alpha > 1$. 

Individual hosts recover at rate $\gamma$ and vectors progress from exposed to infectious at rate $\sigma$. Finally, vectors die at rate $\mu$ and hence their mean lifespan is $1/\mu$. Vector deaths are linked to births of susceptible vectors at the individual level so the number of vectors remains fixed at $M$. The events, corresponding transitions and rates for the within-household model are summarised in Table \ref{tab:rates}.

\begin{table}[ht!]
\begin{tabular}{c|c|c}
\hline
Event & Transition & Rate\\
\hline
Host infection & $(X,Y) \to (X-1, Y+1)$ & $\beta \alpha p X (M-S-E) /N$ \\
Host recovery & $ Y \to Y-1 $ & $\gamma Y$ \\
Vector exposure & $(S,E) \to (S-1, E+1)$ & $\beta \alpha q S Y / N $\\
Vector becomes infectious & $E \to E-1$ & $\sigma E$ \\
Vector death ($E$) & $(S,E) \to (S+1, E-1)$ & $\mu E$\\
Vector death ($I$) & $S \to S+1$ & $\mu(M-S-E)$ \\
\end{tabular}
\caption{Events, transitions and rates that define the within-household CTMC. Note that only the states that change in a given transition are shown, all other states remain the same.}
\label{tab:rates}
\end{table}

In our model, hosts are mobile, but vectors are assumed to stay within their household. Hence between-household transmission occurs via the homogeneous mixing of hosts within the overall population. For everything that follows, we will assume households are all comprised of (fixed) $N$ hosts and (fixed) $M$ vectors, for simplicity, but this could be extended to a distribution over households of different compositions. We assume a large and initially completely susceptible population of households (both hosts and vectors), with the exception of a single household, which allows us to model the initial stages of an outbreak as a branching process \cite{Athreya:1972,Ball:1997}. In general this is a multi-type branching process \cite{Dorman:2004}, but as we show below we can simplify this to just two types for our purposes.

\begin{figure}[ht!]
\begin{center}
\includegraphics[width=0.65\textwidth]{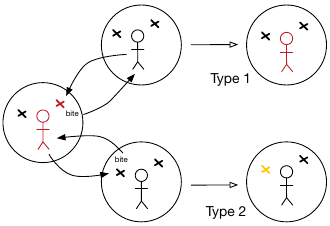}
\caption{\label{fig:model} Between-household transmission dynamics. Black individuals (hosts and vectors) are susceptible, red are infected, and yellow is exposed. Either a susceptible host travels and is bitten by an infected vector, resulting in a Type 1 (a single infected host) configuration, or an infected host moves and is bitten by a susceptible vector resulting in a Type 2 (a single exposed vector) configuration.}
\end{center} 
\end{figure}

As both the infected and susceptible hosts are assumed to mix in the population, between-household transmission can occur via two routes: 
\begin{enumerate}
\item[(i)] a susceptible host travels to a house with infected vectors and is bitten and returns to their house infected; or,
\item[(ii)] an infected host travels to a house with susceptible vectors and is bitten, leading to an exposed vector.
\end{enumerate}
These two transmission routes are illustrated in Figure \ref{fig:model}. The branching process can therefore be formulated using two types corresponding to the two possible initial states of a household infected by the two routes, respectively. For notational convenience we denote these two types as Type 1 (a single infected host) and Type 2 (a single exposed vector). Note that we assume that for the second type, the infected host only potentially infects at most a single vector. This could also be relaxed at the expense of introducing more types in the branching process, an additional type for each possible number of initially-infected vectors, but will not be explored here.
 
\medskip

The rate at which a single infected household creates new Type 1 infected households is 
\begin{equation}
	f_1(\Psi(t)) = \beta (1-\alpha) p (M-S-E),
	\label{eq:f1}
\end{equation}
where $(1-\alpha)$ is the proportion of time a host spends outside their primary household. 
The rate at which a new Type 2 infected household is created is similarly
\begin{equation}
	f_2(\Psi(t)) = \beta(1-\alpha) q \frac{YM}{N}.
	\label{eq:f2}
\end{equation}
Two assumptions should be highlighted: this model only assumes one extra person in each house at any time and does not account for the corresponding reduction in the primary household number of hosts. Both of these can in principle be adjusted for by scaling the transmission rate parameters. 

\section{Epidemiological characteristics}
\label{sec:epichar}

The reproduction dynamics for the households of each type are summarised by the {\em next-generation matrix} \cite{Diekmann2000,Diekmann2009}. As there are two routes of infection into a household, there are two initial conditions we need to consider and hence four reproductive numbers that can be calculated.
The next-generation matrix, $K$, has elements
\begin{equation}
	k_{ij} 
	= 
	\
	\mathbb{E}
	 \left [ 
		\int_0^\infty f_{i}(\Psi(t)_j) dt 
	\right ],
	\quad\quad i,j\in \{1,2 \},
	\label{eq:k_el}
\end{equation}
where $\Psi(0)_j$ is the initial state of a household of Type $j$ and the functions $f_i$, are as given in Eqs.~\eqref{eq:f1} and \eqref{eq:f2}. 
The element $k_{ij}$ is the expected number of households of Type $i$ infected by a single Type $j$ household in a completely susceptible population.
Each of these are path integrals that can be evaluated as detailed in~\cite{PS02,RHK10,BHKR12}, and detailed below.

First note that the state space can be partitioned into a set of absorbing states, $A$, and transient states, $B$, ($\mathbb{S} = A \cup B$). The elements $k_{ij}$ for $j=1,2$, can be evaluated simultaneously for each $i \in \{1,2\}$ by solving a set of linear equations
\begin{equation}
	Q_B x = -f_i,
\end{equation}
where $Q_B$ is the transition rate matrix (i.e.,~the generator) for the within-household CTMC restricted to the transient states, $B$, and $f_i$ is a vector with elements corresponding to the rates in Eqs.~\eqref{eq:f1} and \eqref{eq:f2}. For each $i$, $k_{i1}$ is found by selecting the element of the vector $x$ corresponding to the state with a single infected host ($X=N-1,Y=1,S=M$) and $k_{i2}$, the element corresponding to the state with a single exposed vector ($X=N, S=M-1, E=1$).
We note that the (marginal) {\em offspring distribution} for each type can be calculated in a similar manner by recursively solving a set of linear equations~\cite{RHK10}. 

Once the next-generation matrix, $K$, has been calculated, an overall reproduction number, corresponding to the {\em household basic reproduction number}, $R_*$, for the expected total number of secondary households (regardless of type) can be calculated as the dominant eigenvalue of the next-generation matrix~\cite{Diekmann2000}. The corresponding normalised eigenvector is the {\em proportion of each type} in the (exponentially growing) infected population. 

The {\em early growth rate}, $r$ can be calculated in a similar manner by forming the matrix $G(r)$ with elements
\begin{equation}
 	g_{ij}
 	= 
 	\
	\mathbb{E} \left [ 
		\int_0^\infty f_{i}(\Psi(t)_j) e^{-rt} dt 
	\right ],
	\quad\quad i,j\in \{1,2 \}.
\end{equation} 
For a given value of $r$, each element of $G$ can be evaluated by solving a path integral where the original process, and hence $Q_B$, is modified so that for each transient state an additional rate $r$ is added to the rate of entering the absorbing set $A$ \cite{RHK10}. This is known as exponential discounting. The early growth rate can then be determined by solving 
\begin{equation}
	\Lambda(r) = 1,
\end{equation}
using numerical root finding, where $\Lambda(r)$ is the dominant eigenvalue of the matrix $G(r)$. 

The {\em probability of a major outbreak} is the complement of the probability of extinction. For branching processes, given $T$ types, the $T$ probabilities of extinction starting with a single individual of each type, $p$, satisfies $p = G(p)$ where $G$ is the (joint) probability generating function of the (joint) offspring probability mass function. In~\cite{Ross2015}, for a similar household-structured branching process to that considered here of two main types (but also explicitly handling a distribution of household sizes), it was shown that the probabilities of extinction $p$ satisfied $p = \phi(p)$ where $\phi$ was the (joint) Lapace-Stieltjes transform of the path integrals~\eqref{eq:k_el}, which is the solution of a set of linear equations. We adopt this approach, and report the probability of a major outbreak, $1-p$.

The {\em final size distribution} for hosts within a household can be calculated by considering the hitting probabilities for each possible absorbing state in $A$. The probability of hitting a given set of absorbing states $a \subseteq A$, given the process starts in state $i$, satisfies the system of linear equations~\cite{Nor97},
\begin{equation}
	h_i^a 
	=
	\left\{
	\begin{array}{lr}
	1 \quad & i \in a,\\
	\sum_{j \in \mathbb{S} }\, p_{ij} h_j^a & i \notin a,
	\end{array}
	\right.
\end{equation}
where $p_{ij}$ is the probability of transitioning from state $i$ to $j$, given the system is in state $i$ (i.e.,~an element of the jump chain). The final size distribution can then be determined by solving this system of linear equations for each of the $N$ possible absorbing states corresponding to $Y=0$, $S=M$ and $X=0,\dots,N$. The final size distribution will differ depending on the Type, corresponding to either a single infected host or exposed vector.

MATLAB code to evaluate all these quantities is provided as part of the EpiStruct Project~\cite{Epistruct}.

\section{Results}

We fix some parameters informed by literature, with time units of days, namely: $1/\mu=21$ (mean vector lifespan), $1/ \gamma = 7$ (mean host infectious period), $1/\sigma = 5$ (mean vector exposed period) as are typical of Dengue. We take $p=q=0.7$ (symmetrical probabilities of transmission), $\beta=2\gamma$ (each host is bitten on average 2 times over their infectious period), $\alpha=0.7$ indicating that 70\% of an individual's time is spent in their primary household, and $N=4$ hosts. 

\begin{figure}[ht!]
\begin{center}
\includegraphics[width=0.95\textwidth]{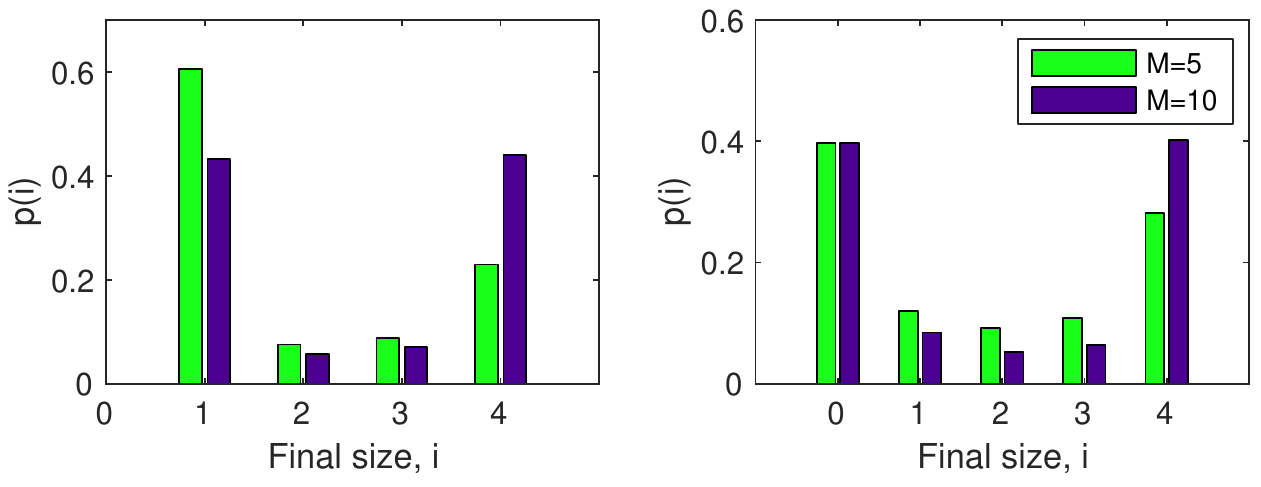}
\caption{Final size distributions for $N=4$ hosts and $M=5$ (green) and $10$ (purple) vectors. Panel (a) shows the distributions for a Type 1 household (a host infected) and (b) a Type 2 household (a vector exposed).}
\label{fig:finalSize1}
\end{center}
\end{figure}

Final size distributions for the number of infected hosts are shown in Figure \ref{fig:finalSize1} for two sizes of vector populations, $M=5$ and $M=10$. Note that the second panel includes zero infections, as the initially infected vector may die before infecting any of the hosts. In both cases we see that the distribution is bimodal indicating with a high probability either a small/no outbreak or the entire household becoming infected as the most-likely outcomes.

\medskip

\begin{figure}[ht!]
\begin{center}
\includegraphics[width=0.95\textwidth]{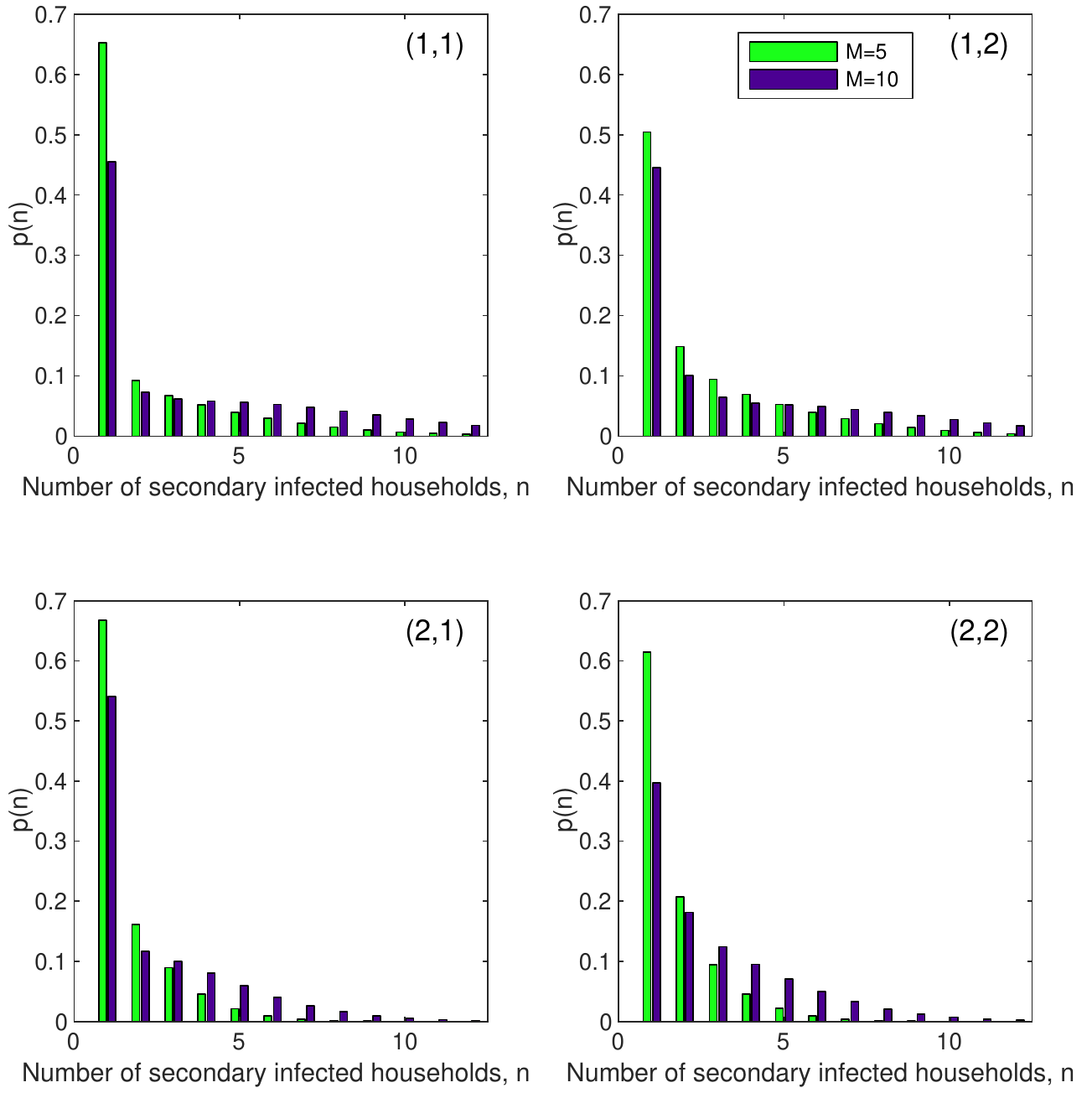}
\caption{Offspring distributions for each combination of types and two numbers of hosts, $M$. Each panel shows the offspring distribution, i.e.~the probability of a single Type $j$ household infecting $n$ Type $i$ households, where the numbers in the upper right corner of each panel indicate $(i,j)$, where Type 1 is infected host and Type 2 is exposed vector; the pattern of the types is therefore the same as in the next generation matrix.}
\label{fig:offspring1}
\end{center}
\end{figure}

The next generation matrices for the two scenarios with $M=5$ and $M=10$ are  
$$
K_5
=
\begin{pmatrix}
    1.21  &  1.67 \\
    0.71  &  0.65 \\
\end{pmatrix},
\quad\text{and}\quad
K_{10}
=
\begin{pmatrix}
    3.11  &  3.02 \\
    1.85  &  1.46 \\
\end{pmatrix},
$$
with corresponding marginal offspring distributions for each type shown in Figure \ref{fig:offspring1}. 
The overall household reproductive numbers are then $R_* = 2.1$ ($M=5$) and $R_* = 4.8$ ($M=10$) respectively, with corresponding household type proportions $(0.66,0.33)$ and $(0.64,0.36)$ -- these are the proportions of each type of household (1, infected host, and 2, exposed vector) in an exponentially growing population of infected households in the branching process approximation~\cite{Diekmann2000}. 

For these parameters we observe that the primary route of transmission is from susceptible hosts visiting and being bitten in infected households, rather than infected hosts travelling and infecting susceptible vectors. In the case of $M=5$ vectors, the probability of a major outbreak starting with a Type 1 (infected host) and Type 2 (exposed vector) household, respectively, is $0.33$ and $0.38$, increasing to $0.60$ and $0.54$, respectively, when there are $M=10$ vectors. The increase in $M$ leads to a corresponding decrease in the doubling time of the epidemic from 25 to 10 days. 

\noindent
{\em Control interventions comparison}

We illustrate the usefulness of this modelling approach by comparing two interventions aimed at reducing virus spread. We assume that spraying, which is the primary measure used to control vector numbers, decreases the number of vectors in each household, $M$. Another intervention available in this model is to increase the proportion of time hosts spend in their (primary) house $\alpha$, which decreases the rate of mixing between households at the expense of spending more time in the primary household. 

\begin{figure}[ht!]
\begin{center}
\includegraphics[width=0.95\textwidth]{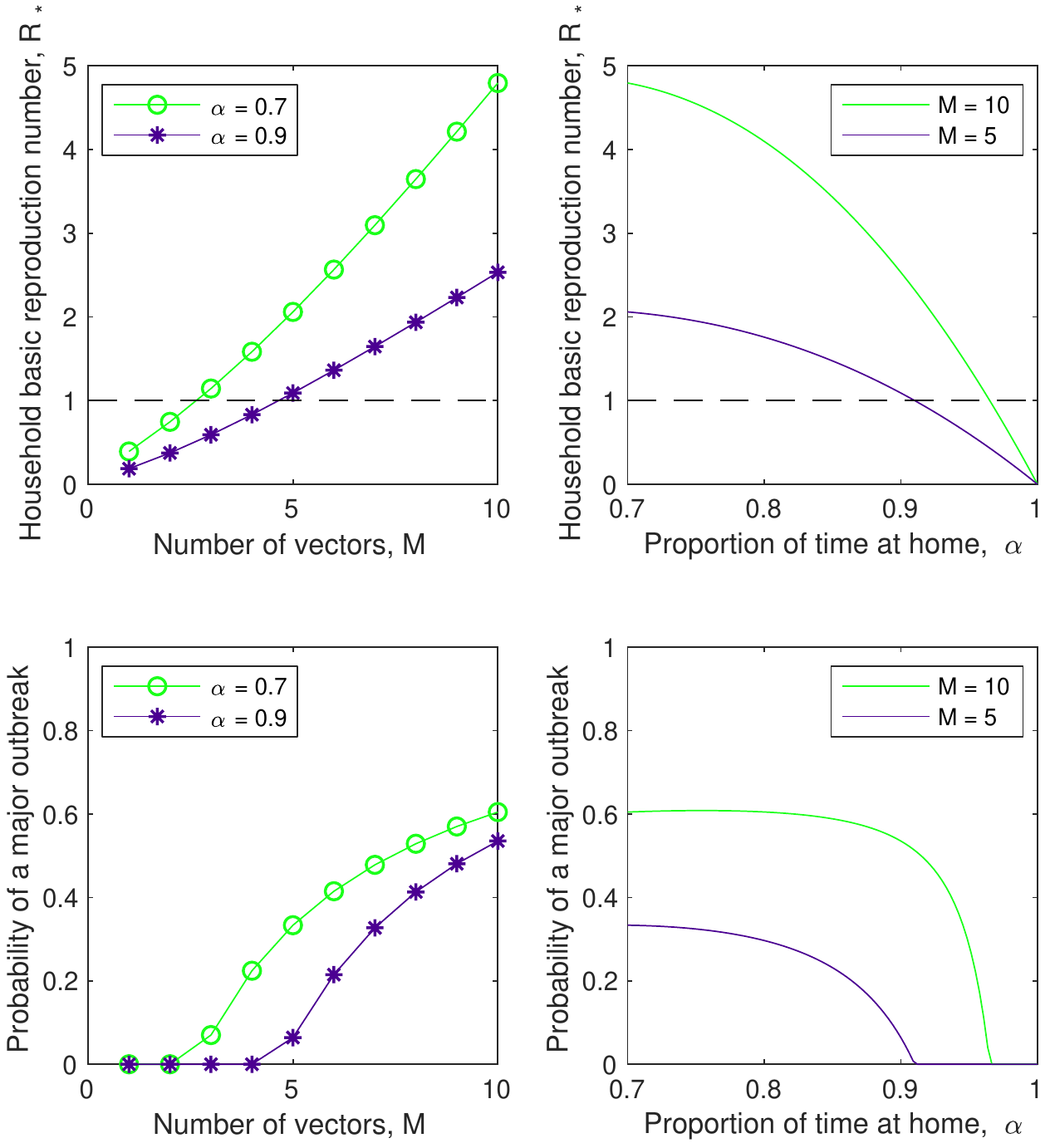}
\caption{Interventions -- by either decreasing the number of vectors within a household, $M$, or increasing the proportion of time spent at home, $\alpha$ -- impact on the household basic reproduction number, $R_*$, and the probability of a major outbreak starting with a Type 1 (infected host) household.}
\label{fig:intervention}
\end{center}
\end{figure}

Figure~\ref{fig:intervention} shows the affect of varying the number of hosts, $M$, and proportion of time spent in the primary household, $\alpha$, on the overall household reproduction number, $R_*$, and the probability of a major outbreak starting with a Type 1 (infected host) household. It can be seen that to almost surely avoid a major outbreak, the number of vectors (per household) $M$ would need to be reduced to less than three when the proportion of time spent in your (primary) household $\alpha=0.7$; however, the number of vectors (per household) $M$ would need to be reduced only to less than five if the proportion of time spent in your (primary) household $\alpha=0.9$. Control via social distancing (increasing $\alpha$) does not appear to be a suitable control mechanism alone; but might be able to be used for a short-period following a potential incursion to provide some synergistic impact to spraying (reducing the number of hosts (per household) $M$). It can be seen that while the household basic reproduction number $R_*$ is relatively sensitive to the proportion of time spent at home ($\alpha$) (at least as $M$ increases), the probability of a major outbreak is not, requiring a substantial increase in the proportion of time spent at home ($\alpha$) to mitigate the risk of a major outbreak in isolation of spraying.

\section{Discussion}

The clustering and limited dispersal of the {\em Ae.~aegypti} mosquito, the primary vector for a number of diseases, has driven a number of recent model developments to understand the role that human movement plays in the dynamics of these vector-borne diseases. Those previous models incorporating explicit household structure and stochasticity have been detailed, simulation-based models. The model we introduce here is much simpler, with the benefit of facilitating analytical tractability and efficient computation of a wide-range of characteristics of epidemiological interest.

We illustrated the utility of this model by comparing two possible control interventions, of spraying to reduce the number of vectors per household, and social distancing to decrease the amount of time spent mixing outside primary households. This highlighted that the probability of a major outbreak is relatively insensitive to social distancing and hence it should not be used as a control measure in isolation; however, some synergistic benefit could be gained when used in conjunction with the more impactful use of spraying.

Obviously the simplicity of the model introduced here, while facilitating efficient computation, means the model has limitations and a greater number of assumptions than the more detailed models mentioned earlier. A significant assumption is that mixing outside of the primary household occurs in the population at random; hence, there is no accounting for the regular patterns that occur in some of these contacts. While undoubtedly a limitation, the impact of this is likely to be less critical in the early stages of an outbreak, which is the stage at which the purpose of the model (as a branching process) is focused. An avenue for further research is to consider three-level mixing models, which might be able to capture some of the regular between-household contacts.

Some of the other limitations of the model we specifically studied herein -- that we did not allow for multiple exposed vectors when an infected host visits a susceptible household, nor for a distribution of household host and vector sizes -- can be accommodated at the expense of introducing more types (rather than only two) and computation. Further, other reproduction numbers can be calculated~\cite{Pellis09}.

\section*{Data Availability}

MATLAB code to evaluate all quantities studied in this paper are provided as part of the EpiStruct Project~\cite{Epistruct}.

\section*{Acknowledgements}
JVR thanks the University of Adelaide for supporting his Special Studies Program in which this work was completed. 

\bibliography{adelaide_refs.bib}

\end{document}